\documentclass{sig-alternate-05-2015}

\usepackage{multirow}
\usepackage{graphicx}
\usepackage{subfig}
\usepackage{hyperref}
\usepackage{xpatch}
\usepackage{url}
\usepackage{fancyhdr}

\makeatletter
\def\@copyrightspace{\relax}
\makeatother

\def\sharedaffiliation{%
\end{tabular}
\begin{tabular}{c}}
\fancypagestyle{footer}{\fancyhf{}\fancyfoot[c]{\textit { \small IWSPA 2017 Proceedings of the 3rd ACM International Workshop on Security and Privacy Analytics, co-located with CODASPY'17, pages 65-72, Scottsdale, Arizona, USA - March 24 - 24, 2017 }}}

\begin{document}

\CopyrightYear{2017}
\setcopyright{acmcopyright}
\conferenceinfo{IWSPA'17,}{March 24 2017, Scottsdale, AZ, USA}
\isbn{978-1-4503-4909-3/17/03}\acmPrice{\$15.00}
\doi{http://dx.doi.org/10.1145/3041008.3041010}
\clubpenalty=10000 
\widowpenalty = 10000






%

\title{{\ttlit EMULATOR} vs {\ttlit REAL PHONE}: Android Malware Detection Using Machine Learning}
%
%
%
%
%

%
\numberofauthors{3}
    \author{
      \alignauthor Mohammed K. Alzaylaee\\      
      \email{malzaylaee01@qub.ac.uk}
      \alignauthor Suleiman Y. Yerima\\     
      \email{s.yerima@qub.ac.uk}
      \alignauthor Sakir Sezer\\    
      \email{s.sezer@qub.ac.uk}
      \sharedaffiliation
      \affaddr{Centre for Secure Information Technologies (CSIT)}  \\
      \affaddr{Queen's University Belfast}   \\
      \affaddr{Belfast, Northern Ireland}
          }
%

\thispagestyle{footer}

%



\maketitle
\begin{abstract}
The Android operating system has become the most popular operating system for smartphones and tablets leading to a rapid rise in malware. Sophisticated Android malware employ detection avoidance techniques in order to hide their malicious activities from analysis tools. These include a wide range of anti-emulator techniques, where the malware programs attempt to hide their malicious activities by detecting the emulator. For this reason, countermeasures against anti-emulation are becoming increasingly important in Android malware detection. Analysis and detection based on real devices can alleviate the problems of anti-emulation as well as improve the effectiveness of dynamic analysis. Hence, in this paper we present an investigation of machine learning based malware detection using dynamic analysis on real devices. A tool is implemented to automatically extract dynamic features from Android phones and through several experiments, a comparative analysis of emulator based vs. device based detection by means of several machine learning algorithms is undertaken. Our study shows that several features could be extracted more effectively from the on-device dynamic analysis compared to emulators. It was also found that approximately 24\% more apps were successfully analysed on the phone. Furthermore, all of the studied machine learning based detection performed better when applied to features extracted from the on-device dynamic analysis.
\end{abstract}

%
%

%
%

%
%

\section{Introduction}
The Google Android operating system (OS) is the leading OS in the market with nearly 80\% market share compared to iOS, Blackberry, Windows, and Symbian mobile. Over 1 billion Android devices have been sold with an estimated 65 billion app downloads from Google Play  ~\cite{president}. Moreover, it is reported that more than 1.5 billion Android devices will be shipped worldwide by 2020 ~\cite{statista}. This has led to malware developers increasingly targeting Android devices. According to a report from McAfee, there are more than 12 million Android malware samples with nearly 2.5 million new Android malware samples discovered every year ~\cite{McAfeeLabs2016}.

The rapid increase in malware numbers targeting Android devices has highlighted the need for efficient detection mechanisms to detect zero-day malware. In contrast with other mobile operating systems, Android allows users to download applications from third party stores, many of which do not have any mechanisms or tools to check the submitted apps for malware. The Google play store uses a tool called Bouncer to screen submitted applications. However it has been previously demonstrated that the Bouncer dynamic analysis process can be bypassed by means of some simple anti-emulator techniques ~\cite{Oberheide2012}.

Android malware can be found in a variety of applications such as banking apps, gaming apps, media player apps etc. These malware-infected apps may access phone data to collect privacy sensitive information, root the phone, dial premium rate numbers or send text messages to premium rate numbers without the user approval etc. Many Android malware families employ detection avoidance techniques in order to hide their malicious activities and evade anti-virus software. Commonly used detection avoidance methods by Android malware include a wide range of anti-emulator techniques, where the malware programs attempt to hide their malicious activities when being analysed in an emulator. For this reason, countermeasures against anti-emulation are becoming increasingly important in Android malware detection.

Several approaches for anti-emulation (or anti-virtualization) have been discussed in previous work. The paper ~\cite{Vidas2014} discusses some methods that could be employed in order to detect the run-time environment thereby hindering dynamic analysis. Some malware applications, for example, will detect the emulator through the use of Android APIs. For instance, if the Telephony Manager API method TelephonyManager.getDeviceId() returns 000000000000000, it means the run-time environment is an emulator rather than a real device, because no real phone will return 0s as the device identifier. This is one of the emulator detection methods used by the Pincer family of Android malware ~\cite{AndroidPincer}.

The emulator can also be detected through the networking environment which is different from that of a real phone, or the underlying QEMU can be detected. Morpheus  ~\cite{Jing2014} has also exposed more than 10,000 detection heuristics based on some artifacts that can be used to detect the run-time analysis environments. These artifacts can be used in malware samples to hide the malicious activities accordingly. Dynamic analysis tools that rely on emulators (or virtual devices) such as Dynalog ~\cite{Alzaylaee2016} attempt to address the problem by changing properties of the environment to emulate a real phone as much as possible and to incorporate several behaviours to mimic a real phone. However, these methods whilst useful, have been shown to be insufficient to completely tackle anti-emulation  \cite{Irolla2016}, \cite{Jing2014}, ~\cite{DroidDolphin} .

Several dynamic analysis tools such as TaintDroid ~\cite{Enck2010}, DroidBox ~\cite{DroidBox}, CopperDroid ~\cite{Tam2015}, Andrubis ~\cite{ANDRUBIS}, AppsPlayground ~\cite{AppsPlayground} have been proposed. Similarly, some online based tools are available for Android malware analysis such as SandDroid ~\cite{SandDroid}, CopperDroid ~\cite{copperdroid}, TraceDroid ~\cite{tracedroid}, and NVISO ApkScan ~\cite{nviso}. All of these dynamic approaches can be evaded by malware that use anti-emulation. Since the analysis is done in a virtualized environment.

Mutti et al. ~\cite{Mutti2015}, have demonstrated the feasibility of device-based dynamic analysis to alleviate the problems of anti-emulation. We have also found that phones generally provide more effective analysis environment due to incomplete nature of emulators. Many apps nowadays have functionality that utilize device hardware features such as sensors, WiFi, GPS, etc. and these require emulation in the sandbox. Therefore, devices should provide more effective analysis environment. Hence, we have designed and implemented a python-based tool to enable dynamic analysis using real phones to automatically extract dynamic features and potentially mitigate anti-emulation detection. Furthermore, in order to validate this approach, we undertake a comparative analysis of emulator vs device based detection by means of several machine learning algorithms. We examine the performance of these algorithms in both environments after investigating the effectiveness of obtaining the run-time features within both environments. The experiments were performed using 1222 malware samples from the Android malware genome project ~\cite{malgenomeproject} and 1222 benign samples from Intel Security (McAfee Labs).

The rest of the paper is structured as follows. Section II describes the runtime analysis process for feature extraction from the phone and emulator, Section III details the methodology and experiments undertaken for the evaluation. The results and the discussions of results will be given in Section IV, followed by related work in Section V. Section VI will present conclusions and future work.
\section{Phone Based Dynamic Analysis and Feature Extraction}
In order to apply machine learning to the classification and detection of malicious applications, a platform is needed to extract features from the applications. These features will be used by the machine learning algorithm to classify the application. Prior to that, the algorithm must be trained with several instances of clean and malicious applications. This process is known as supervised learning. Since our aim is to perform experiments to compare emulator based detection with device based detection we need to extract features for the supervised learning from both environments. For the emulator based learning, we utilized the DynaLog dynamic analysis framework described in ~\cite{Alzaylaee2016}.

The framework is designed to automatically accept a large number of applications, launch them serially in an emulator, log several dynamic behaviours (features) and extract them for further processing. DynaLog components include an emulator-based analysis sandbox, an APK instrumentation module, Behaviour/features logging and extraction, App trigger/exerciser and log parsing and processing scripts ~\cite{Alzaylaee2016}. DynaLog provides the ability to instrument each application with the necessary API calls to be monitored, logged and extracted from the emulator during the run-time analysis. The instrumentation module was built using APIMonitor ~\cite{APIMonitor}. 

\begin{figure}[b]
\centering
\includegraphics[height=.5in, width=3.3in]{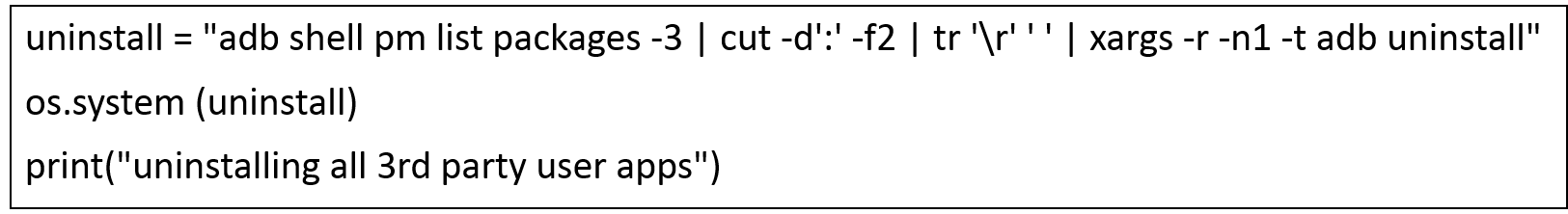}
\caption{Uninstalling third party apps from the phone using package manager in an adb (android debug bridge) shell.}
\label{uninstall1}
\end{figure}

DynaLog currently relies on Monkey, an application exerciser tool, which allows the application to be triggered with thousands of random events for this experiment. These random events include "swipes", "presses", "touch screens" etc, to ensure that most of the activities has been traversed to cover as much code as possible. In order to enable dynamic analysis and feature extraction from a real phone, the DynaLog framework was extended with a python-based tool to enable the following:
\begin{itemize}
\item	At the start, push a list of contacts to the device SD card and then import them (using adb shell command) to populate the phone's contact list.
  \item Discover and uninstall all third-party applications prior to installing the app under analysis (or from which features are to be extracted). This capability was implemented using package manager within an adb shell as illustrated in Fig. \ref{uninstall1}.
  \item Check whether the phone is in airplane mode or not. If it is in airplane mode, turn it (airplane mode) off. This is because with airplane mode switched on, many phone features such as WiFi, 3G/4G connectivity will be unavailable, which could affect the dynamic analysis. Also, Monkey (the app exerciser tool) was found to sometimes temper with the analysis by turning on the airplane mode due to its randomly sending events such as "touch screens", "presses" and "swipes".
  \item Check the battery level of the phone. If the level is very low i.e. battery has discharged to a low level, put the analysis on hold (sleep) until the phone has re-charged to an appropriate level.
	\item	Outgoing call dialling using adb shell.
	\item	Outgoing sms messages using adb shell.
	\item	Populate the phone SD card with other assets such as folders containing dummy files: image files, pdf, text files etc. 
\end{itemize}

\begin{figure}[t]
\centering
\includegraphics[height=1.238in, width=3.36in]{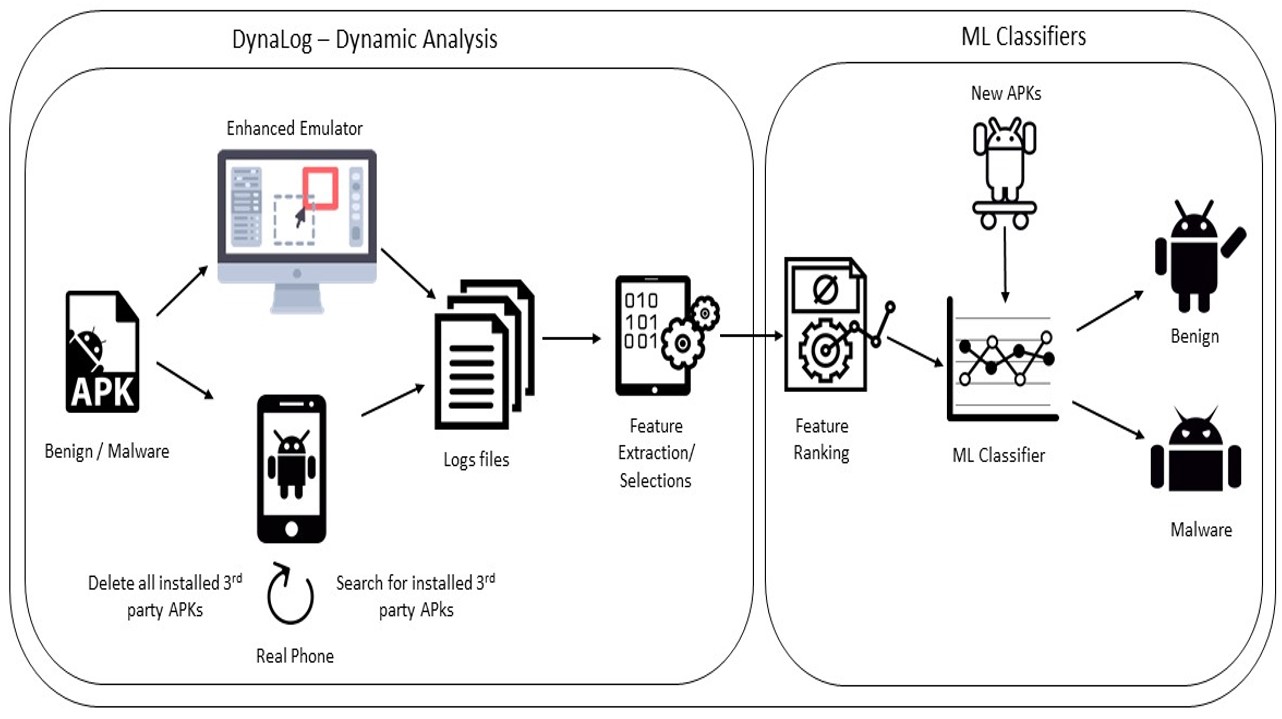}
\caption{Phone and emulator based feature extraction using DynaLog.}
\label{architecture2}
\end{figure}

Each of the apps is installed and run for 300 seconds on the emulator (and then on the phone for the second experiment) and the behaviours are logged and parsed through scripts that extract the desired features. The features used in the experiments include API calls and Intents (signalling critical events). The API calls signatures provided during the instrumentation phase allows the APIs to be logged and extracted from the phone (or the emulator) via adb logcat as the app is run. For malicious apps that incorporate anti-emulator techniques it is expected that the API call that could lead to the exposure of their malicious behaviour would not be logged as the malware will attempt to hide this behaviour. The analysis process is shown in Fig. \ref{architecture2}.

\section{METHODOLOGY AND EXPERIMENTS}
This section describes the methodology of the experiments which were performed to evaluate the environmental effectiveness of detecting and extracting the features as well as analysing the performance of the machine learning algorithms on both emulator and real phone. 
\subsection{Dataset}
The dataset used for the experiments consists of a total of 2444 Android applications. Of these, 1222 were malware samples obtained from 49 families of the Android malware genome project ~\cite{malgenomeproject}. The rest were 1222 benign samples obtained from Intel Security (McAfee Labs). 
\subsection{Environmental configurations}
The two environments under which the experiments were performed had the following configurations. The first was the phone which was an Elephone P6000 brand smartphone equipped with Android 5.0 Lollipop, 1.3 GHz CPU, 16 GB internal memory, and 32 GB of external SD card storage. In addition, a sim card with call credit was installed in the phone to enable outgoing calls and 3G data usage.

The emulator environment (an Android virtual device) was created in a Santoku Linux VirtualBox based machine. The emulator was configured with 2 MB of memory, and an 8 GB of external SD card memory. The Android version in the emulator was 4.1.2 Jelly Bean (API level 16). Note that, the features are app dependent so the difference in Android versions will have no impact. The emulator was enhanced as described in ~\cite{Alzaylaee2016} by adding contact numbers, images, .pdf files, and .txt files as well as changing the default IMEI, IMSI, sim serial number and phone numbers to mimic a real phone as much as possible. The emulator is also restored after each application run to ensure that all third party apps are removed. 
\subsection{Features extraction}
After all the apps are run and processed in both analyses environments, the outputs are pre-processed into a file of feature vectors representing the features extracted from each application. Within the vector, each feature is represented by 0 or 1 denoting presence or absence of the feature. The files were converted to ARFF format and fed into WEKA machine learning tool for further processing. Initially, 178 features were extracted for both the phone and emulator environments. These were then ranked using the InfoGain (information gain) feature ranking algorithm in WEKA. The top 100 ranked features were then used for the experiments to compare the detection performance between the two environments using several machine learning algorithms. The top 10 ranked features (based on InfoGain) in both sets (phone and emulator) are shown in Table \ref{InfoGainRanking}. 
\begin{table}[t]
\small
\renewcommand{\arraystretch}{1}
\caption{Top-10 Ranked Features Based on InfoGain}
\label{InfoGainRanking}
\centering
\begin{tabular}{c||p{3.5cm}||p{3.5cm}}
\hline
\centering \textbf{} & \centering {\textbf{Top-10 Features from Emulator}} & {\centering {\textbf{Top-10 Features from Phone}}} \\ \hline
\textbf{1} & BOOT\_COMPLETED & BOOT\_COMPLETED \\ \hline
\textbf{2} & Ljava/util/Timer;\-schedule & UMS\_DISCONNECTED \\ \hline
\textbf{3} & Ljava/io/FileOutputStream ;-write & Ljava/lang/ClassLoader;\-getResourceAsStream \\ \hline
\textbf{4} & UMS\_DISCONNECTED & Ljava/util/Timer;\-schedule \\ \hline
\textbf{5} & Ljava/lang/ClassLoader;\-getResourceAsStream & Ljava/lang/ClassLoader;\-getResource \\ \hline
\textbf{6} & Ljava/lang/ClassLoader;\-getResource & INSTALL\_REFERRER \\ \hline
\textbf{7} & Landroid/content/pm/ ApplicationInfo;\-getApplicationInfo & Ljava/lang/Class;\-getClassLoader \\ \hline
\textbf{8} & INSTALL\_REFERRER & Ljava/io/FileOutputStream ;-write \\ \hline
\textbf{9} & Ljava/lang/Class;\-getClassLoader & Ljava/lang/ClassLoader \\ \hline
\textbf{10} & Ljava/lang/ClassLoader & Ljava/lang/System;\-loadLibrary \\ \hline
\end{tabular}
\end{table}
\subsection{Machine learning classifiers}
In the experiments, the features were divided into five different sets (20, 40, 60, 80 and 100 top ranked using InfoGain) in order to compare the performance of emulator and phone based detection using machine learning algorithms. The following algorithms were used in the experiments:  Support Vector Machine (SVM-linear), Naive Bayes (NB), Simple Logistic (SL), Multilayer Perceptron (MLP), Partial Decision Trees (PART), Random Forest (RF), and J48 Decision Tree.
\subsection{Metrics}
Five metrics were used for the performance emulation of the detection approaches. These include: true positive rate (TPR), true negative ratio (TNR), false positive ratio (FPR), false negative ratio (FNR), and weighted average F-measure. The definition of these metrics are as follows:
\begin{equation}
\label{TR}
TPR = \frac{TP}{TP+FN}
\end{equation}
\begin{equation}
\label{TNR}
TNR = \frac{TN}{TP+FN}
\end{equation}
\begin{equation}
\label{FPR}
FPR = \frac{FP}{TP+FN}
\end{equation}
\begin{equation}
\label{FNR}
FNR = \frac{FN}{TP+FN}
\end{equation}
\begin{equation}
\label{F-measures}
F-measure = \frac{2* recall * precision}{recall + precision}
\end{equation}

True positives (TP) is defined as the number of malware samples that are correctly classified, whereas the false negatives is defined as the number of malware samples that are incorrectly classified. True negatives (TN) is defined as the number of benign samples that are correctly classified, while false positives (FP) is defined as the number of benign samples that are incorrectly classified. The F-measures is the accuracy metric which incorporates both the recall and the precision.
\section{Results and Discussions}
\subsection{Experiment 1: Emulator vs Device analysis and feature extraction}
In order to validate our phone-based machine learning detection, we analysed the effectiveness of running the samples and extracting features from both phone and emulator environments. Out of the 1222 malware samples used, 1205 were successfully run on a real phone compared to only 939 successfully run on an emulator. From the benign samples, 1097 out of 1222 were successfully examined using a real phone versus 786 from the emulator for the same dataset. Therefore, 98.6\% malware sample were successfully analysed using the phone, while only 76.84\% were successfully analysed when using the emulator. Also, 90\% of the benign samples were successfully analysed using the phone versus 64.27\% using the emulator. That gives us a total of 94.3\% successfully analysed using the phone compared to only 70.5\% using the emulator as illustrated in Table \ref{Percentages} . 
\begin{table}[b]
\centering
\caption{Percentage of successfully analysed Android apps}
\label{Percentages}
\begin{tabular}{|c|c|c|}
\hline
                         & \textbf{Emulator} & \textbf{Phone} \\ \hline
\textbf{Malware samples} & 76.84\%           & 98.6\%         \\ \hline
\textbf{Benign samples}  & 64.27\%           & 90\%           \\ \hline
\textbf{Total}           & 70.5\%            & 94.3\%         \\ \hline
\end{tabular}
\end{table}

The phone-based experiments were performed using a Santoku Linux-based VM. During the study we discovered that the use of USB 2.0 or 3.0 was essential. In our initial experiments where the default USB 1.0 was utilized to attach the phone, only 480 out of the 1222 benign samples were able to run. This was due to the fact that samples > 1 MB in size took longer to analyse and hence experienced a 'time-out' error with the use of the slower USB 1.0 standard. We also noticed that apps that did not have any Activities crashed and could not be analysed on the phone or emulator. This accounted for a small percentage of the apps that failed to run successfully. 

\begin{figure}[b]
\centering
\includegraphics[width=3.3in,height=5cm]{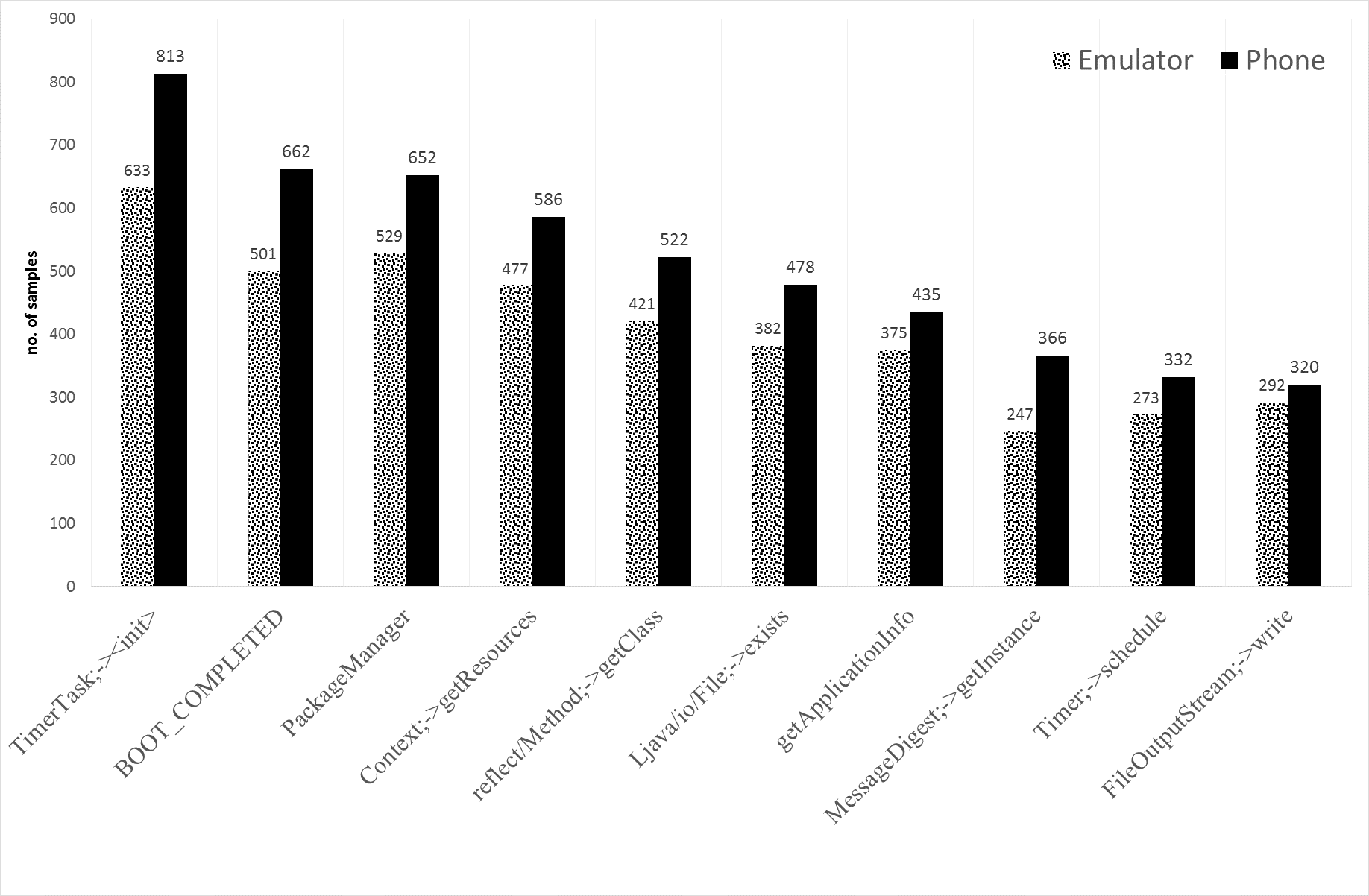}
\caption{Top-10 features extracted from the malware dataset in two different analysis environments (Emulator and Phone-USB2.0)}
\label{malware dataset}
\end{figure}

\begin{figure}[t]
\centering
\includegraphics[width=3.3in,height=5cm]{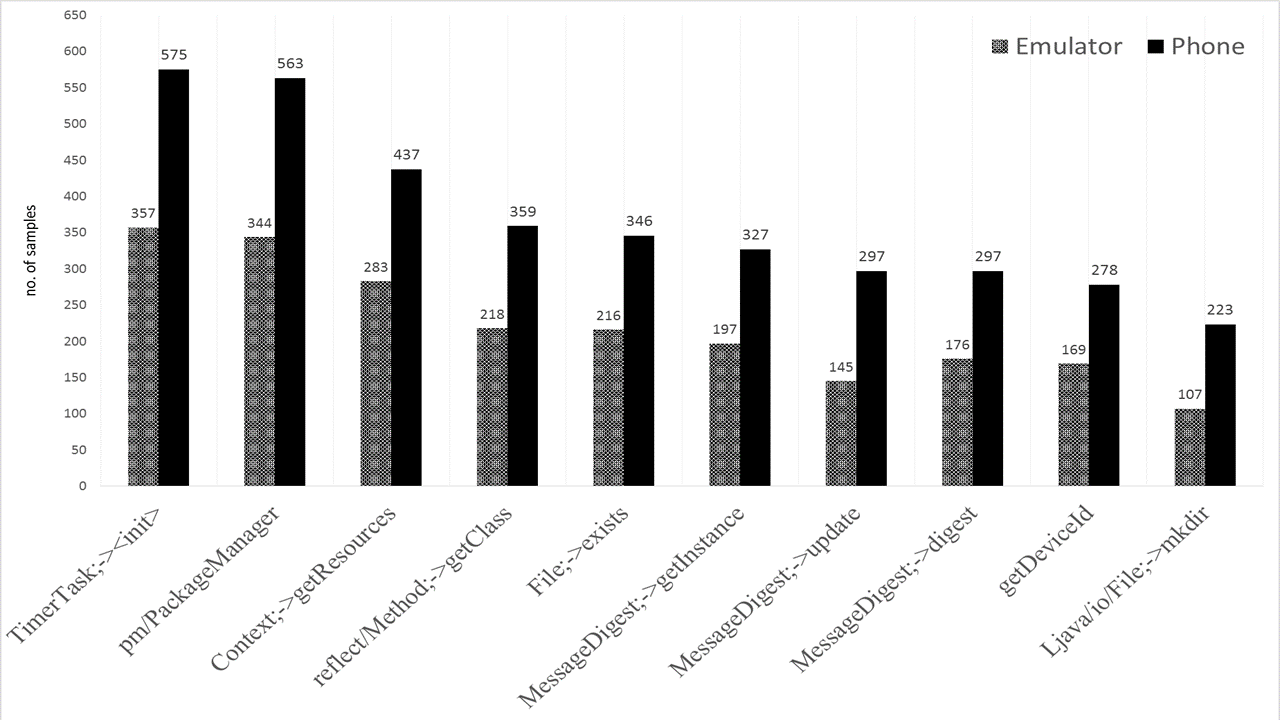}
\caption{Top-10 features extracted from the benign dataset in two different environments (Emulator and Phone-USB2.0)}
\label{benign dataset}
\end{figure}

Fig. \ref{malware dataset} and Fig. \ref{benign dataset} show the top-10 extracted features from malware and benign dataset respectively. Fig. \ref{malware dataset} shows that more malware features are able to be extracted from the phone based analysis vs. the emulator based analysis from the same sample set. The feature "TimerTask;-><init>", for example, was logged from 813 malware applications using the phone, while it was only logged from 633 malware applications using the emulator. Similarly, the feature "intent.BOOT\_COMPLETED" in Fig. \ref{malware dataset}, has been extracted from 662 malware applications using the phone whereas only 501 were extracted from the same sample set using the emulator.

Similar findings appear with the benign samples as shown in  Fig. \ref{benign dataset}. More samples were able to be analysed using the phone. With some features, the difference between phone and emulator extraction were > 200. The more features we can extract during the dynamic analysis the better the result of the detection mechanism is likely to be. There were some features that were extracted exclusively from the phone but not with the emulator. These are shown in Table  \ref{extracted}. The System.loadLibrary feature (found in 209 apps) is the API call associated with native code. The reason it is not logged when the emulator is used could be due to the lack of ability to run native code on the emulator. Overall, the phone based analysis shows a much higher efficiency of detecting and extracting features for analysis of the apps or training machine learning classifiers.
\begin{table}
\centering
\small
\caption{Features extracted from Phone but not obtained with the emulator}
\label{extracted}
\begin{tabular}{|p{4.9cm}|c|c|}
\hline
\centering
\textbf{Extracted Feature}   & \textbf{Phone} & \textbf{Emulator} \\ \hline
Ljava/lang/System;->loadLibrary        & 209            & 0                 \\ \hline
Ljava/net/URLConnection;->connect     & 14             & 0                 \\ \hline
Landroid/content/Context;->unbindService      & 4              & 0                 \\ \hline
Landroid/app/Service;->onCreate           & 3              & 0                 \\ \hline
BATTERY\_LOW                  & 1              & 0                 \\ \hline
Landroid/telephony/SmsManager;->sendTextMessage & 1              & 0                 \\ \hline
\end{tabular}
\end{table}
\subsection{Experiment 2: Emulator vs Device Machine learning detection comparison}
Table \ref{Emulator} and Table \ref{Phone} present the performance evaluation results of the different machine learning algorithms (for the top 100 features). The results demonstrates that using the phone-based features for the dynamic analysis approach is more effective for detecting and classifying the applications compared to the emulator. The results shown were obtained from testing on 33\% of the samples while 66\% were used for training the model. Table 5 shows that higher detection rates were obtained from the phone-based features for all the algorithms (top 100 training features). TPR of > 90\% were obtained with all but the N.B classifier with the phone-based features. Whereas, NB, SL, PART, and J48 all had < 90\% TPR with the emulator-based features.

As mentioned earlier, 939/1222 malware and 786/1222 benign samples executed successfully on the emulator (i.e. total of 1725 out of 2444 samples). Out of these 1725 samples, 12 did not execute successfully on the phone. Thus there were 1713 (939 malware and 774 benign) overlapping apps that executed successfully on both the emulator and the phone. Another set of experiments were performed with only the apps that executed in BOTH the emulator and phone successfully. The results (for top 100 features) using 66\% training and 33\% testing split are shown in Tables \ref{Emulator-BOTH} and \ref{Phone-BOTH} respectively. 

The results show with the exception of RF, all of the tested algorithms in the data collected from emulator recorded an F-measure of < 0.9, whereas with the phone, only NB, PART and J48 recorded < 0.9 F-measure. Again, the results indicate better performance with the phone-based features.

\begin{figure*}
\centering
\subfloat{\includegraphics[width=2.3in,height=3.1cm]{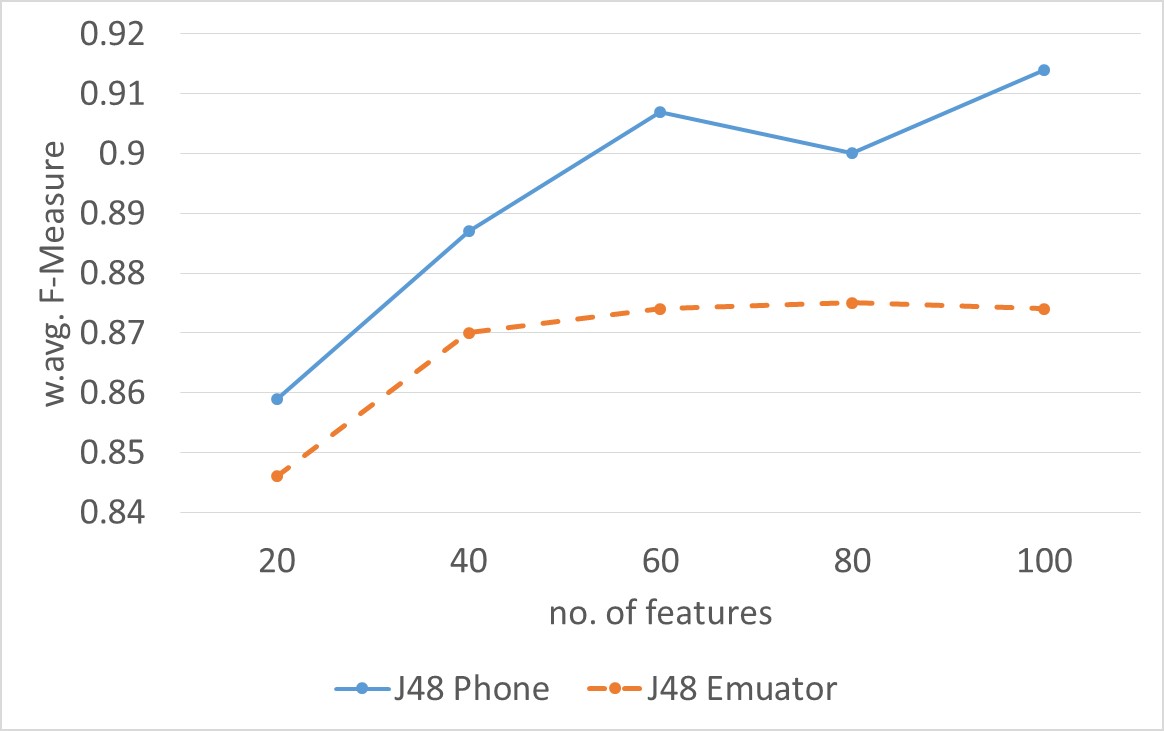}%
\label{J48}}
\hfil
\subfloat{\includegraphics[width=2.3in,height=3.1cm]{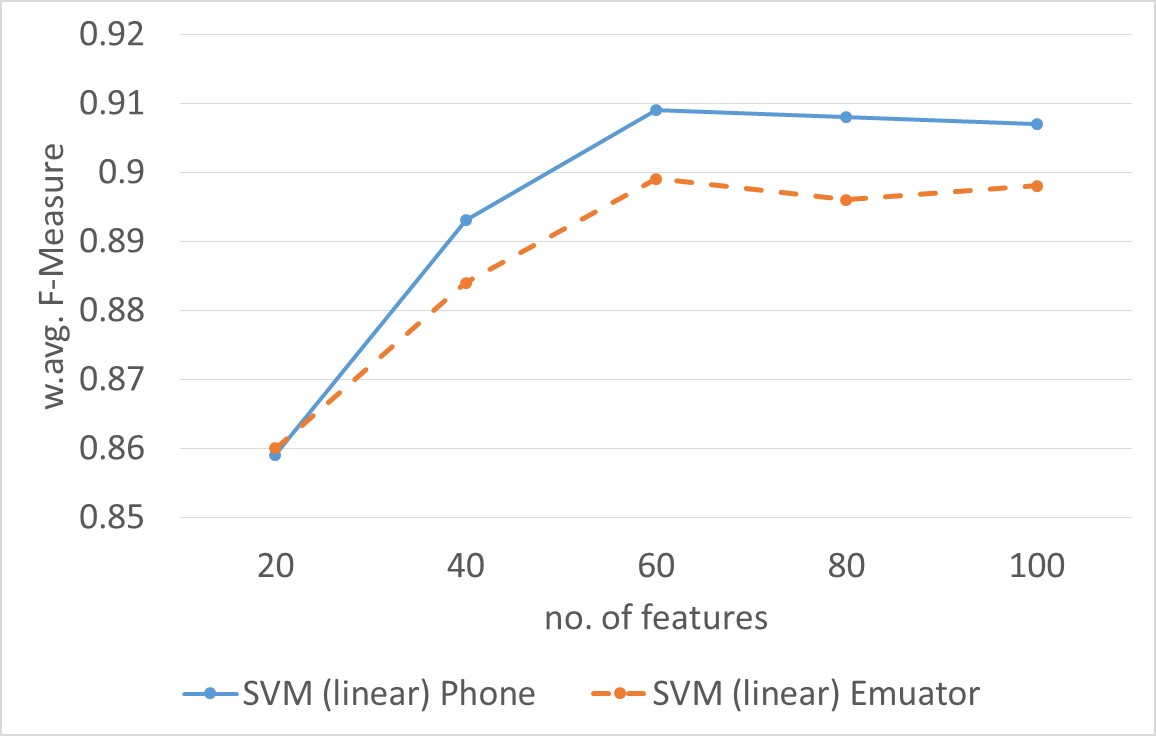}%
\label{SVM}}
\hfil
\subfloat{\includegraphics[width=2.3in,height=3.1cm]{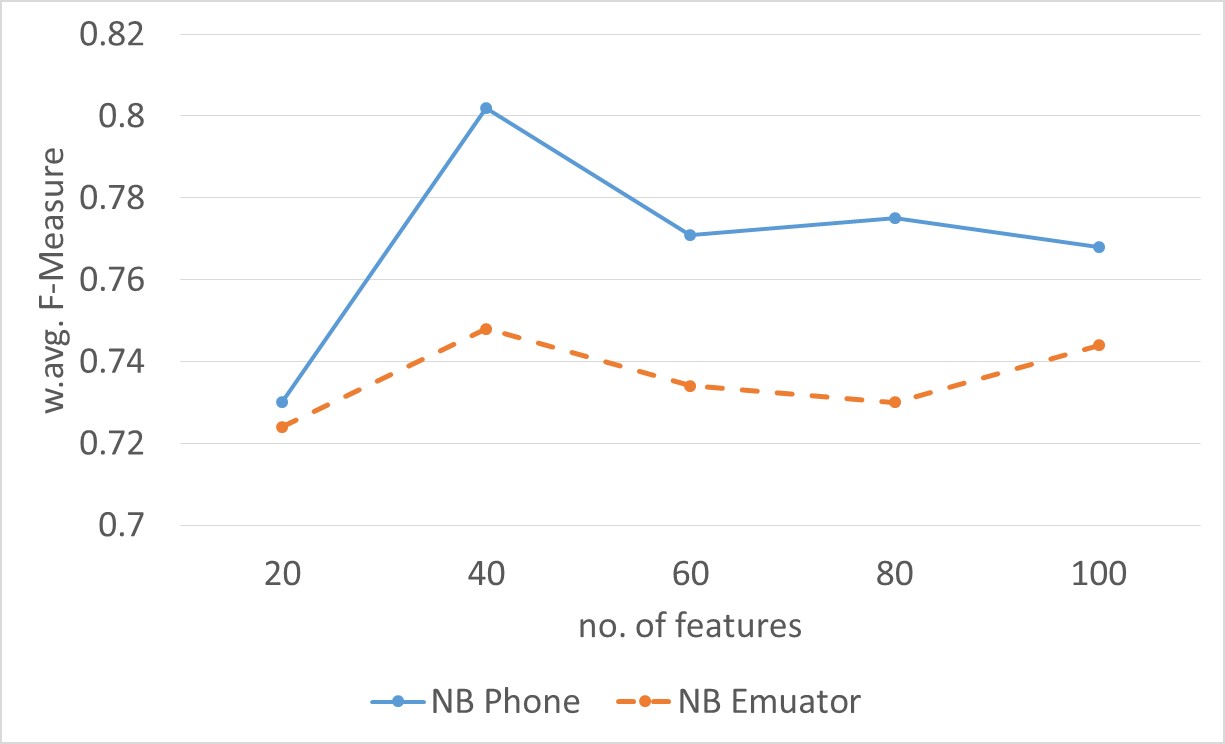}%
\label{NB}}
\hfil
\subfloat{\includegraphics[width=2.3in,height=3.1cm]{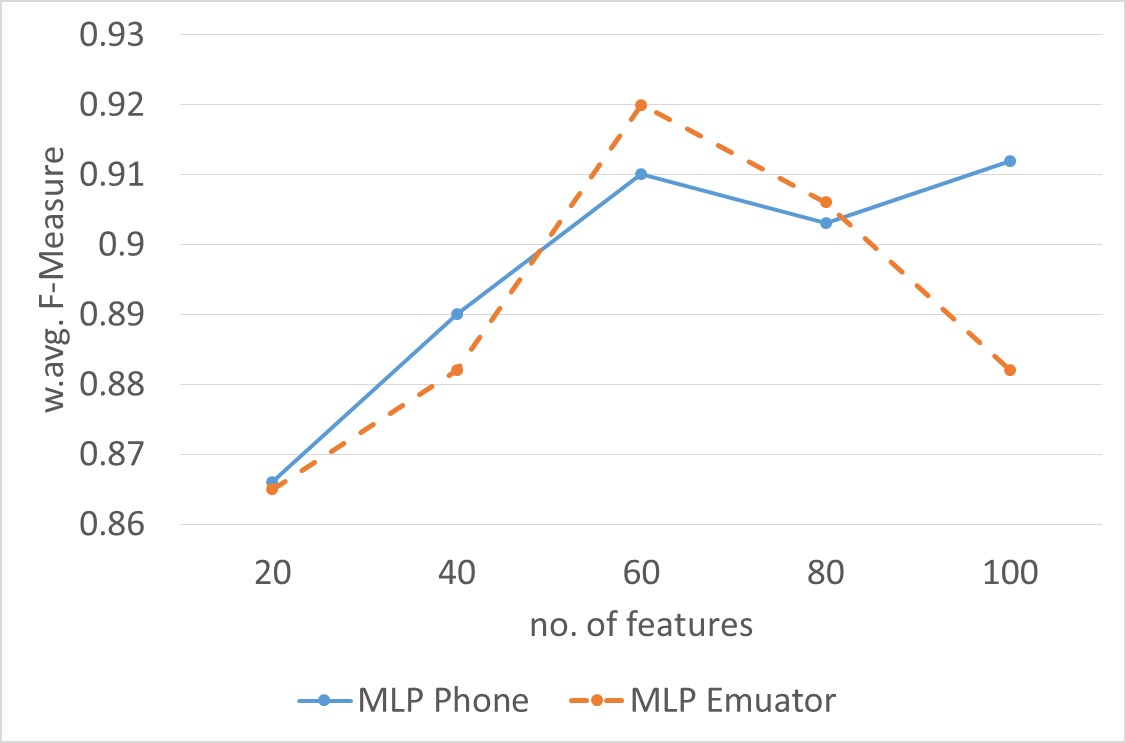}%
\label{MLP}}
\hfil
\subfloat{\includegraphics[width=2.3in,height=3.1cm]{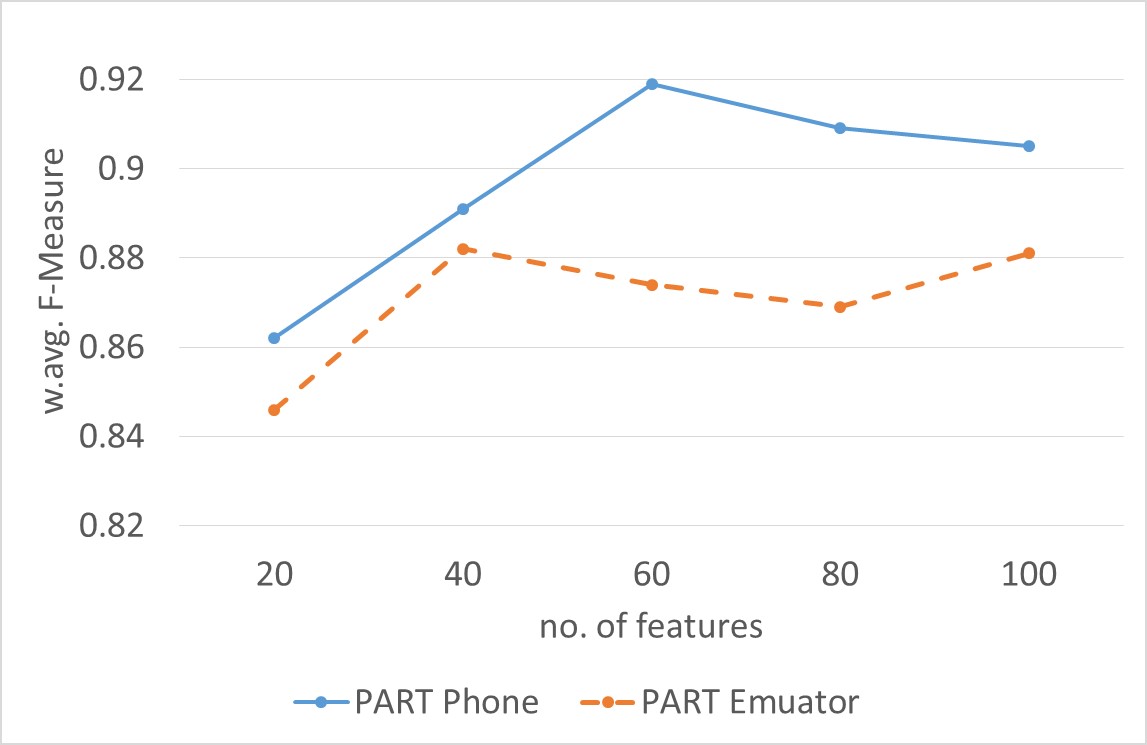}%
\label{PART}}
\hfil
\subfloat{\includegraphics[width=2.3in,height=3.1cm]{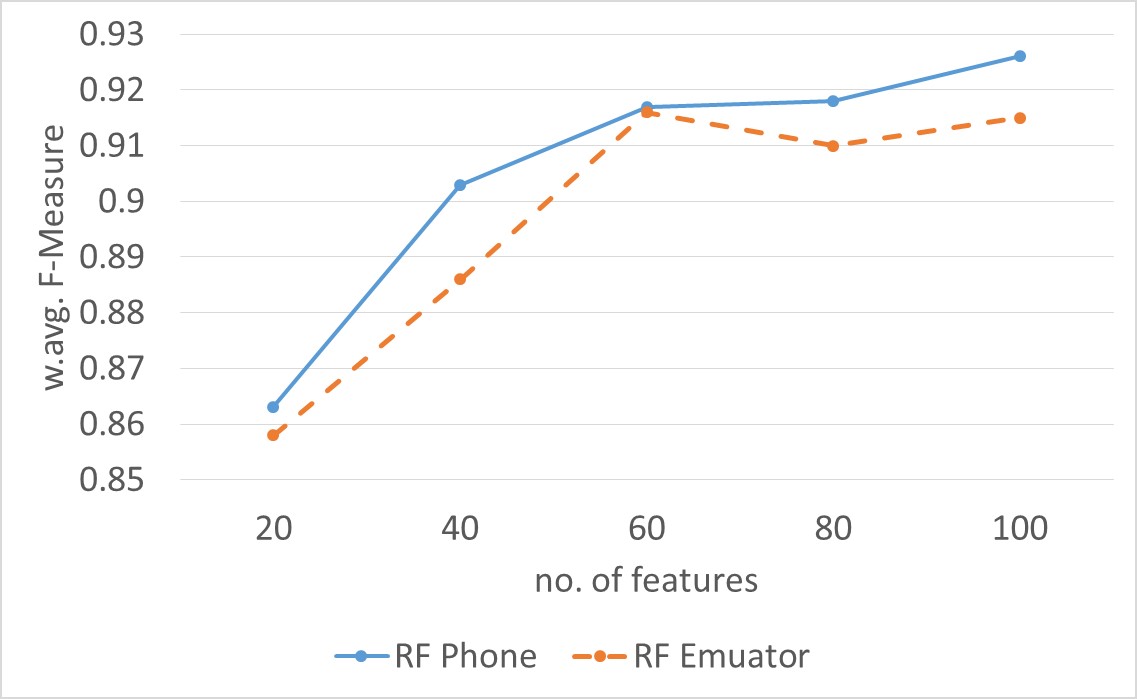}%
\label{RF}}
\hfil
\subfloat{\includegraphics[width=2.3in,height=3.1cm]{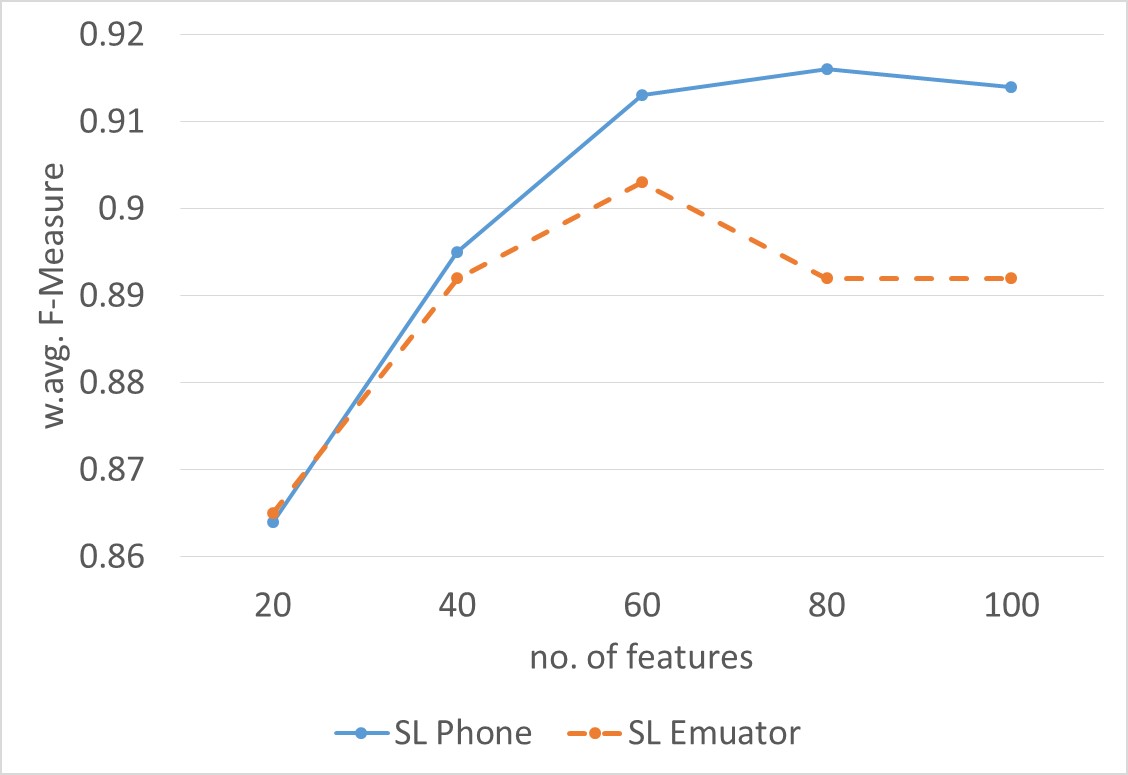}%
\label{SL}}
\caption{F-measure for top 20,40,60,80 and 100 features. Phone vs. emulator comparison.}
\label{fig_Overall}
\end{figure*}

Fig. \ref{fig_Overall} shows the results from top 20, 40, 60, 80 and 100 information-gain ranked features for the first experiment. It depicts the F-measures obtained for the various machine learning algorithms trained on both phone and emulator data. From the figure it is clear that the overall detection performance is better with the phone than the emulator. It is only in the case of MLP-80 features that the emulator records a better f-measure performance. 

\begin{figure*}
\centering
\subfloat{\includegraphics[width=2.3in,height=3.1cm]{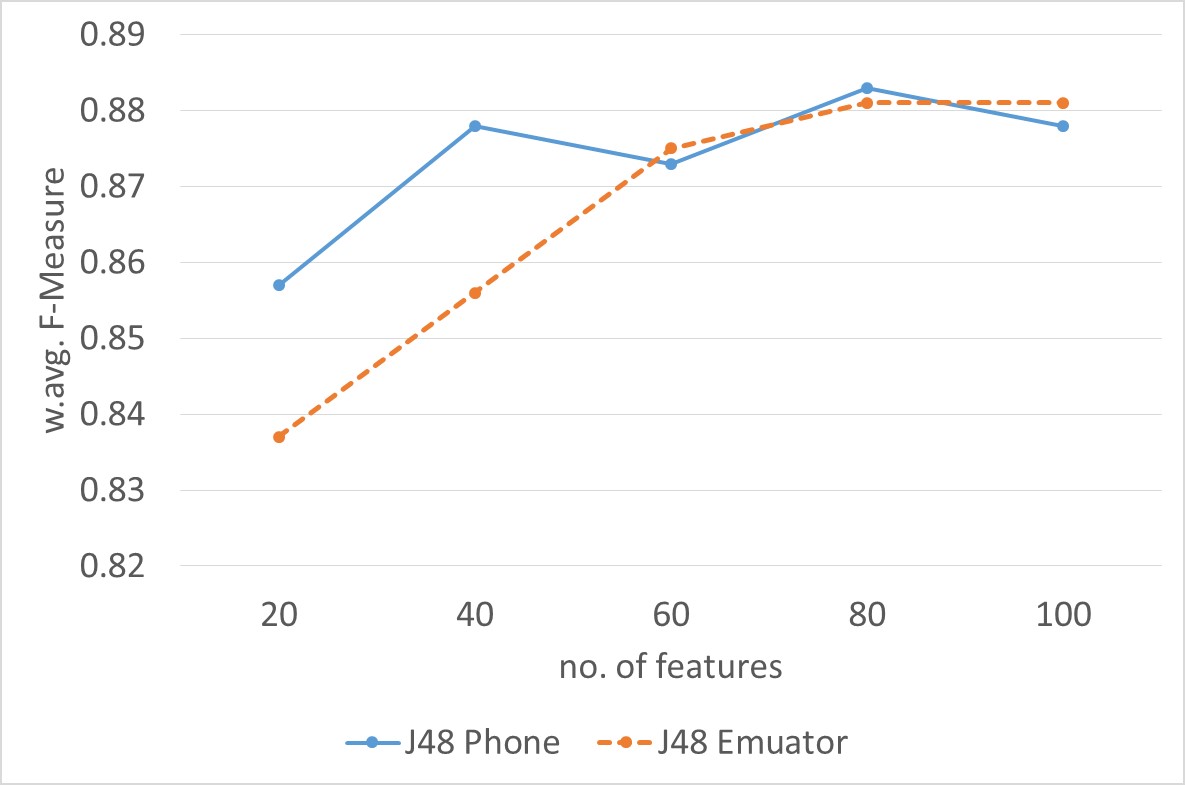}%
\label{J48}}
\hfil
\subfloat{\includegraphics[width=2.3in,height=3.1cm]{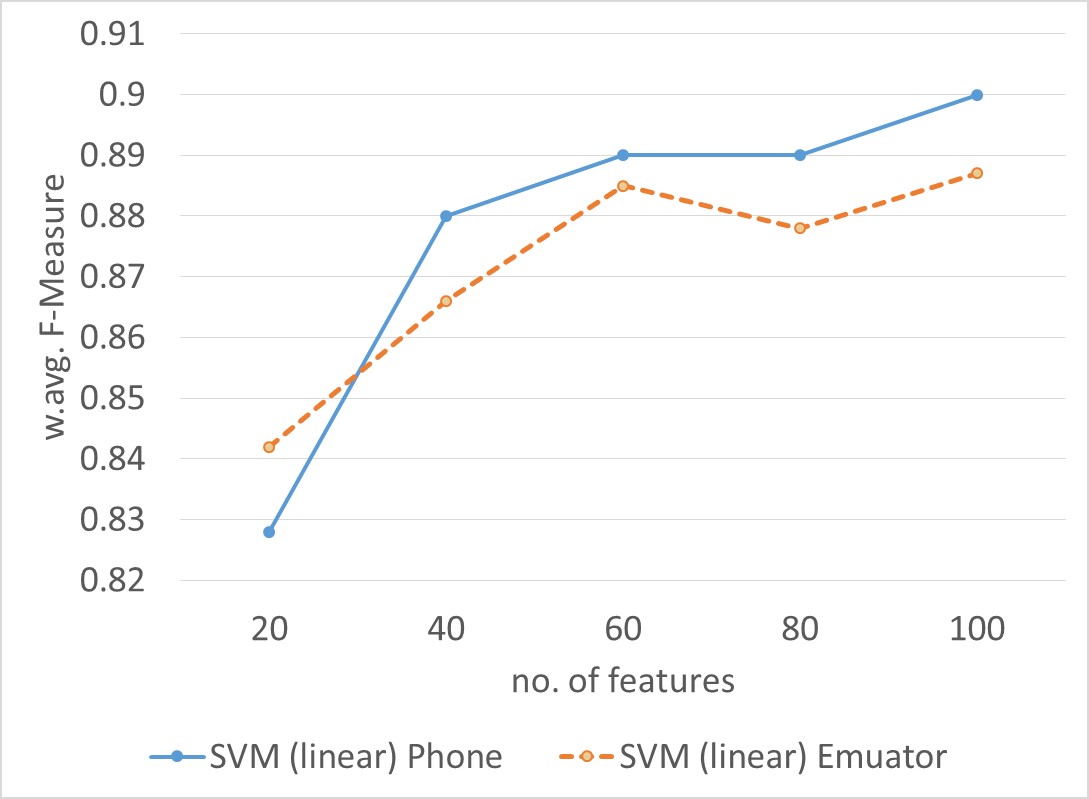}%
\label{SVM}}
\hfil
\subfloat{\includegraphics[width=2.3in,height=3.1cm]{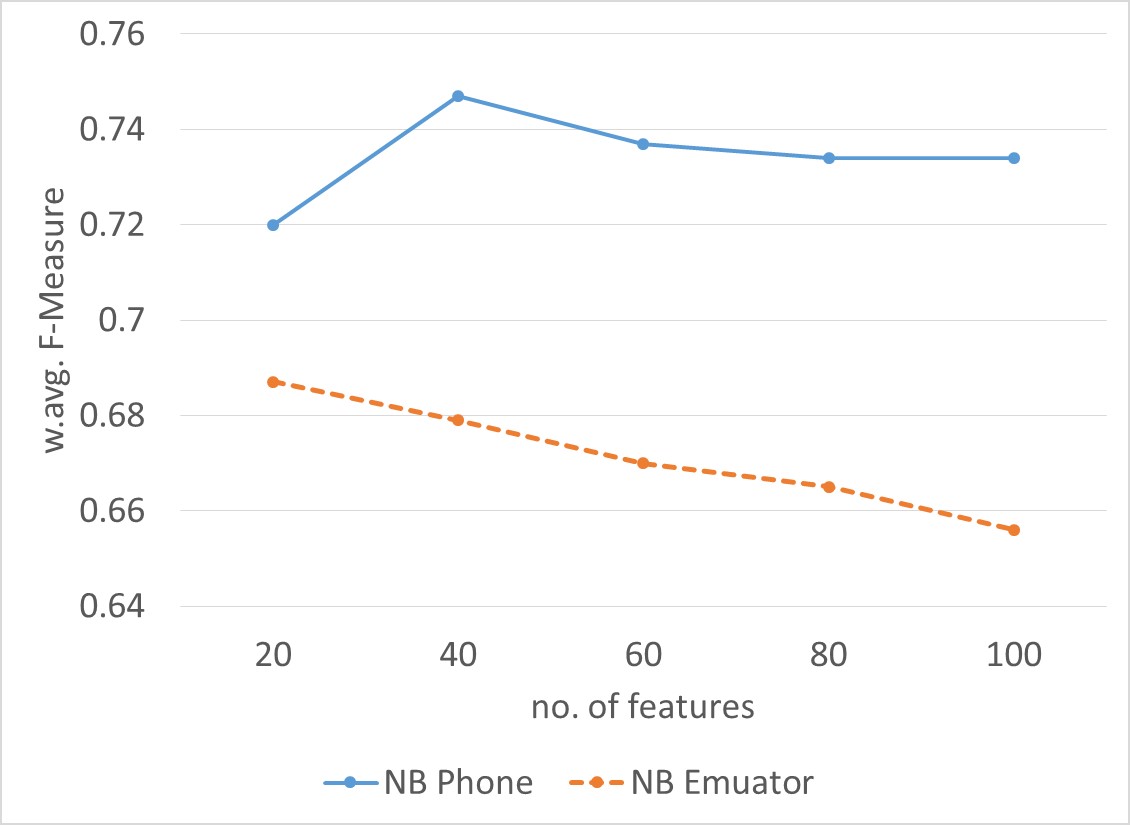}%
\label{NB}}
\hfil
\subfloat{\includegraphics[width=2.3in,height=3.1cm]{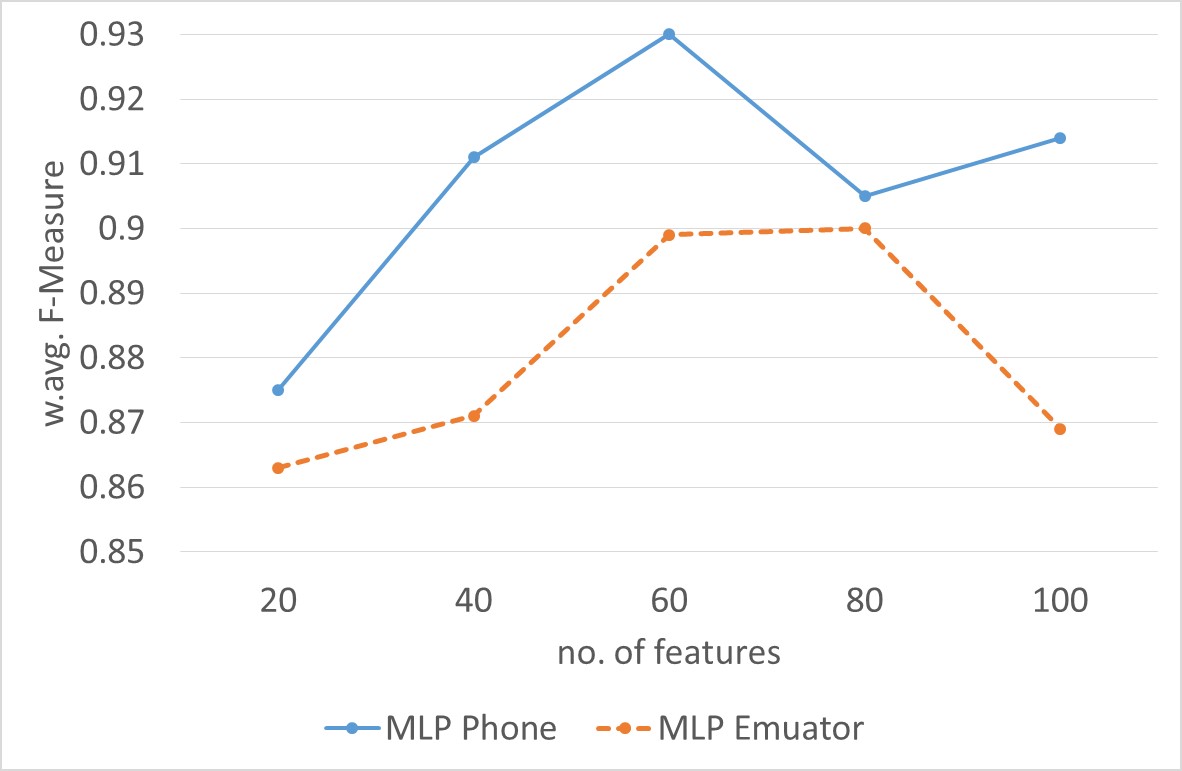}%
\label{MLP}}
\hfil
\subfloat{\includegraphics[width=2.3in,height=3.1cm]{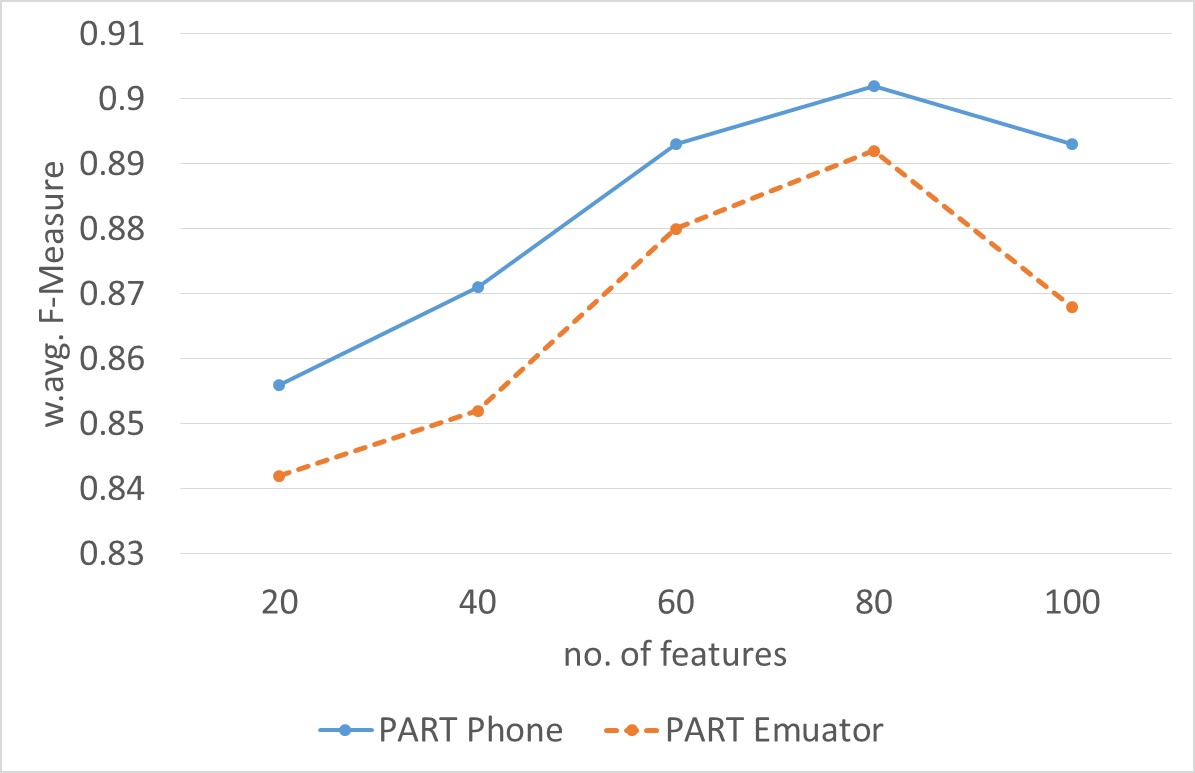}%
\label{PART}}
\hfil
\subfloat{\includegraphics[width=2.3in,height=3.1cm]{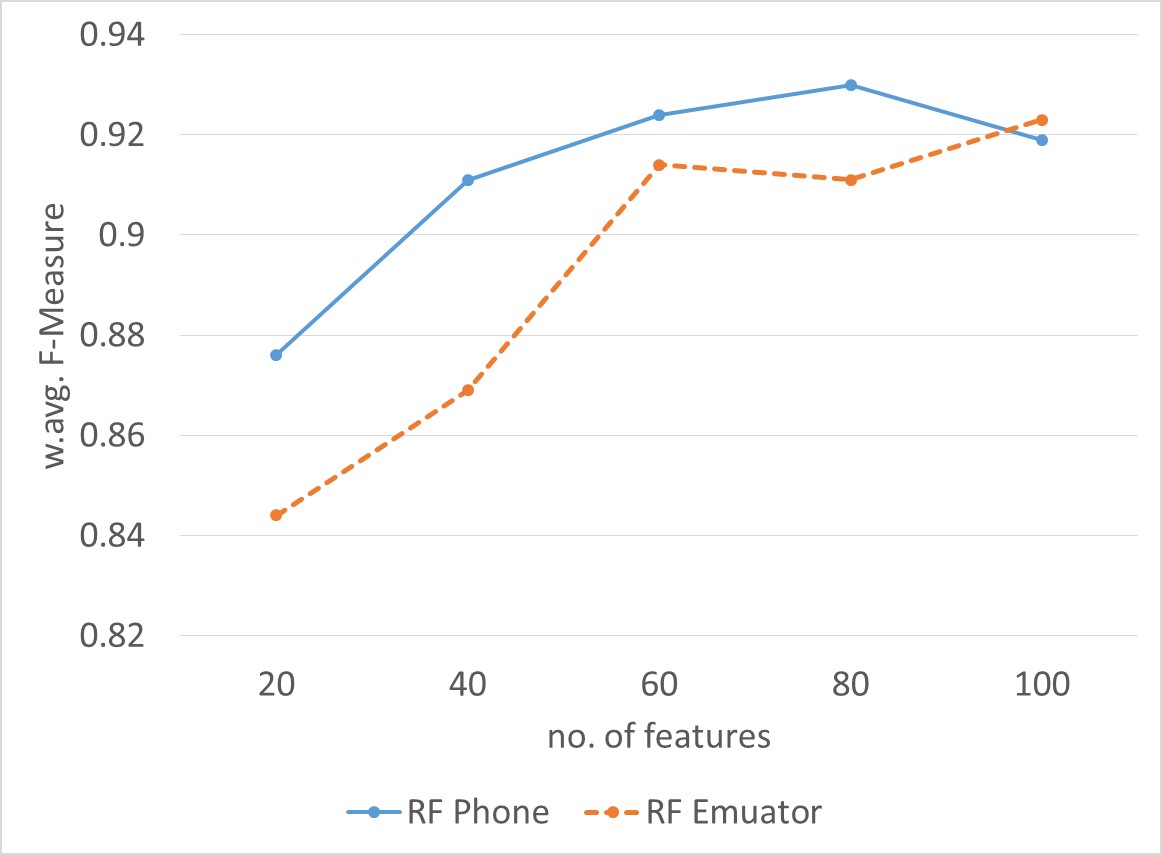}%
\label{RF}}
\hfil
\subfloat{\includegraphics[width=2.3in,height=3.1cm]{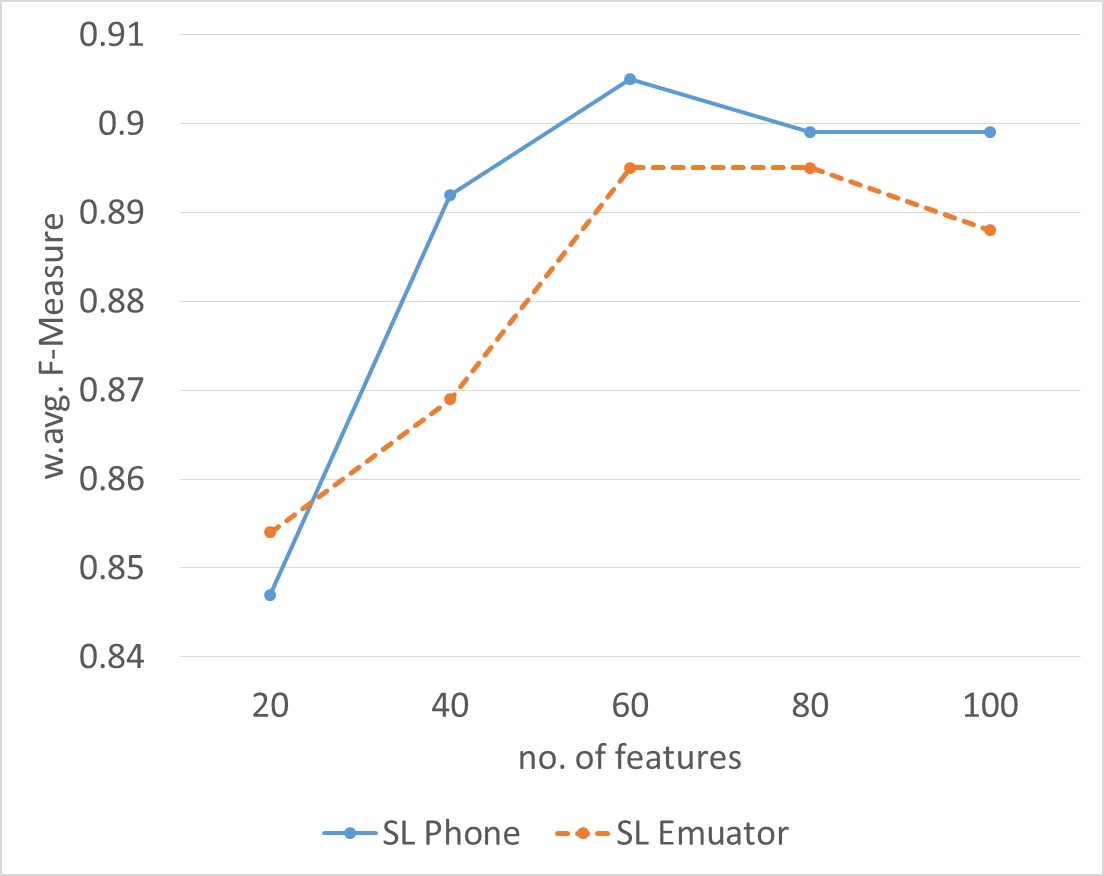}%
\label{SL}}
\caption{F-measure for top 20,40,60,80 and 100 features. Phone vs. emulator comparison for training with samples that executed successfully on BOTH emulator and phone.}
\label{fig_same}
\end{figure*}

Fig. \ref{fig_same} shows the results from top 20, 40, 60, 80 and 100 information-gain ranked features for the second experiment, where only apps that executed successfully in BOTH environments were used. In all the cases, the phone-based experiments showed better performance except in the following: J48-60 features, SVM-20 features, RF-100 features and SL-20 features. Thus, we can conclude that the phone-based method still surpassed the emulator-based method in overall performance prediction.
\begin{table}[!h]
\renewcommand{\arraystretch}{1.1}
\caption{Performance Evaluation of the machine learning algorithms trained from Emulator-based features (top 100 features)}
\label{Emulator}
\centering
\small
\begin{tabular}{|c||c||c||c||c||c|}
\hline
\bfseries ML & \bfseries TPR & \bfseries FPR & \bfseries TNR & \bfseries FNR & \bfseries W-FM \\
\hline\hline 				
SVM-linear & 0.909 & 0.109 & 0.891 & 0.091 & 0.9 \\
\hline 				
NB & 0.596 & 0.102 & 0.898 & 0.404 & 0.73 \\
\hline 				
SL & 0.899 & 0.102 & 0.898 & 0.101 & 0.899 \\
\hline 				
MLP & 0.924 & 0.098 & 0.902 & 0.076 & 0.914 \\
\hline 				
PART & 0.896 & 0.109 & 0.891 & 0.104 & 0.894 \\
\hline 			
RF & 0.909 & 	0.068 & 0.932 & 0.091 & 0.919\\
\hline 				
J48 & 0.88 & 0.125 & 0.875 & 0.12 & 0.878 \\
\hline
\end{tabular}
\end{table}

\begin{table}[!h]
\renewcommand{\arraystretch}{1.1}
\caption{Performance Evaluation of the machine learning algorithms from Phone-based features (top 100 features)}
\label{Phone}
\centering
\small

\begin{tabular}{|c||c||c||c||c||c|}
\hline
\bfseries ML & \bfseries TPR & \bfseries FPR & \bfseries TNR & \bfseries FNR & \bfseries W-FM \\
\hline\hline 				
SVM-linear & 0.916 & 0.096 & 0.904 & 0.084 & 0.91 \\
\hline 				
NB & 0.629 & 0.125 & 0.875 & 0.371 & 0.744 \\
\hline 				
SL & 0.919 & 0.085 & 0.915 & 0.081 & 0.917 \\
\hline 				
MLP & 0.919 & 0.088 & 0.912 & 0.081 & 0.916 \\
\hline 				
PART & 0.904 & 0.101 & 0.899 & 0.096 & 0.902 \\
\hline 			
RF & 0.931 & 	0.08 & 0.92 & 0.069 & 0.926\\
\hline 				
J48 & 0.926 & 0.104 & 0.896 & 0.074 & 0.912 \\
\hline
\end{tabular}
\end{table}

\begin{table}[!h]
\renewcommand{\arraystretch}{1.1}
\caption{Evaluation of the machine learning algorithms on Emulator-based features extracted from apps run successfully in BOTH environments (top 100 features)}
\label{Emulator-BOTH}
\centering
\small
\begin{tabular}{|c||c||c||c||c||c|}
\hline
\bfseries ML & \bfseries TPR & \bfseries FPR & \bfseries TNR & \bfseries FNR & \bfseries W-FM \\
\hline\hline 				
SVM-linear & 0.89 & 0.122 & 0.878 & 0.11 & 0.885 \\
\hline 				
NB & 0.537 & 0.18 & 0.82 & 0.463 & 0.654 \\
\hline 				
SL & 0.884 & 0.11 & 0.89 & 0.116 & 0.887 \\
\hline 				
MLP & 0.887 & 0.106 & 0.894 & 0.113 & 0.89 \\
\hline 				
PART & 0.893 & 0.122 & 0.878 & 0.107 & 0.887 \\
\hline 			
RF & 0.911 & 0.069 & 0.931 & 0.089 & 0.919 \\
\hline 				
J48 & 0.869 & 0.094 & 0.906 & 0.131 & 0.885 \\
\hline
\end{tabular}
\end{table}

\begin{table}[!h]
\renewcommand{\arraystretch}{1.1}
\caption{Evaluation of the machine learning algorithms on Phone-based features extracted from apps run successfully in BOTH environments (top 100 features)}
\label{Phone-BOTH}
\centering
\small

\begin{tabular}{|c||c||c||c||c||c|}
\hline
\bfseries ML & \bfseries TPR & \bfseries FPR & \bfseries TNR & \bfseries FNR & \bfseries W-FM \\
\hline\hline 				
SVM-linear & 0.905 & 0.109 & 0.891 & 0.095 & 0.907 \\
\hline 				
NB & 0.596 & 0.102 & 0.898 & 0.404 & 0.73 \\
\hline 				
SL & 0.902 & 0.098 & 0.902 & 0.098 & 0.902 \\
\hline 				
MLP & 0.905 & 0.072 & 0.928 & 0.095 & 0.916 \\
\hline 				
PART & 0.899 & 0.106 & 0.894 & 0.101 & 0.897 \\
\hline 			
RF & 0.918 & 0.064 & 0.936 & 0.082 & 0.926\\
\hline 				
J48 & 0.88 & 0.125 & 0.875 & 0.12 & 0.878 \\
\hline
\end{tabular}
\end{table}

\subsection{Comparison with other works}
In this subsection, we compare our results with those obtained from the DroidDolphin ~\cite{DroidDolphin} and STREAM  ~\cite{STREAM} dynamic analysis frameworks. DroidDolphin is a dynamic analysis framework for detecting Android malware applications which leverages the technologies of GUI-based testing, big data analysis, and machine learning. STREAM is also a dynamic analysis framework based on Andromaly which enables rapid large-scale validation of mobile malware machine learning classifiers. DroidDolphin used 1000 to 64000 balanced malware and benign Android applications. 

In the STREAM approach, the testing set used 24 benign and 23 malware applications while the training set consisted of 408 benign and 1330 malware applications. Both used split training/testing set (T.S.) and 10 fold cross-validation (C.V.) methods. Table \ref{PhonevsSTREAM} shows the comparison between our phone results (100 features) and the STREAM results, while Table \ref{DroidDolphine} shows the results obtained from DroidDolphin.

From the DroidDolphin results, it is obvious that the detection accuracy is increasing as the number of the training samples are increased. The accuracy rate starts from 83\% for the training set with 1000 applications and increased gradually to 92.50\% with 64k applications. Table \ref{PhonevsSTREAM} shows that despite the difference in the testing set numbers between our work and STREAM, our phone based RF, SL, J48 and MLP perform significantly better for T.S. accuracy. In the case of C.V. accuracy, S.L performs better with our phone results, while the RF, J48, and MLP results were close to those of STREAM.  The C.V. accuracy of RF, SL, J48 and MLP from our phone results showed better performance than all the DroidDolphin C.V. results. 

For the T.S. results, our phone SL, J48 and MLP were better than DroidDolphin T.S. results except for the 32k/32k training/testing split T.S. results. The T.S. results from our phone based RF experiments showed better accuracy than all of the DroidDolphin T.S. results. Therefore, based on the encouraging results, we would continue our analysis using the real phones with larger numbers of training samples in future work.

\begin{table}
\centering
\caption{Accuracy results of Phone (100 features) vs STREAM}
\label{PhonevsSTREAM}
\small
\renewcommand{\arraystretch}{1.1}
\begin{tabular}{|c|c|c|c|c|c|}
\hline
\multirow{2}{*}{\textbf{ML }} & \multicolumn{2}{c|}{\textbf{Real Phone}} & \multicolumn{3}{c|}{\textbf{STREAM}}                                      \\ \cline{2-6} 
                                        & \textbf{T.S.}        & \textbf{C.V.}        & \textbf{T.S.}     & \multicolumn{2}{c|}{\textbf{C.V.}} \\ \hline
{\textbf{NB}}                                     & 75.2                        & 72.9                             & 78.91                    & \multicolumn{2}{c|}{79.79}                     \\ \hline
{\textbf{Bayes net}}                               & 74.8                        & 74.1                             & 81.25                    & \multicolumn{2}{c|}{86.23}                     \\ \hline
{\textbf{RF}}                                      & 92.6                        & 92.9                             & 70.31                    & \multicolumn{2}{c|}{94.53}                     \\ \hline
{\textbf{SL}}                                      & 91.4                        & 90.4                             & 68.75                    & \multicolumn{2}{c|}{89.52}                     \\ \hline
{\textbf{J48}}                                     & 91.4                        & 91.1                               & 73.44                    & \multicolumn{2}{c|}{93.43}                     \\ \hline
{\textbf{MLP}}                                     & 91.2                        & 91.2                             & 70.31                    & \multicolumn{2}{c|}{93.91}                     \\ \hline
\end{tabular}
\end{table}

\begin{table}
\renewcommand{\arraystretch}{1}
\caption{Evaluation Results of DroidDolphin (Accuracy)}
\label{DroidDolphine}
\centering
\small
\begin{tabular}{|c||c||c|}
\hline
\bfseries Quantity & \bfseries Testing Set & \bfseries Cross Validation \\
\hline\hline 				
0.5k0.5k & 	83 & 79.90 \\
\hline 				
1k1k & 84 & 81.70 \\
\hline 				
2k2k & 85.80 & 83.30 \\
\hline 				
4k4k & 87.20 & 83.80 \\
\hline 				
8k8k & 89.10 & 83.50 \\
\hline 			
16k16k & 91.30 & 84.10 \\
\hline 				
32k32k & 92.50 & 86.10 \\
\hline
\end{tabular}
\end{table}

\section{RELATED WORK}
Once a new malware application is discovered in the wild, it should be run in a closed environment in order to understand its behaviour. Researchers and malware analysts rely heavily on emulators or virtual devices due to the fact that it is a comparatively low cost analysis environment. Emulators are also more attractive for automated mass analysis commonly employed with machine learning. Hence, most previous machine learning based detection with dynamic analysis rely on feature extraction using tools running on emulator environments. Contrary to previous machine learning based dynamic detection work, we attempt to utilize real phones (devices) for automated feature extraction in order to avoid the problem of anti-emulator techniques being employed by Android malware to evade detection.

Some previous machine learning based Android malware detection works such as \cite{Drebin}, , \cite{Yerima}, \cite{DroidAPIMiner}, \cite{Eigenspace}, have considered API calls and Intents in their studies. However, unlike our work, these are based on static feature extraction and thus could be affected by obfuscation. Marvin \cite{Marvin} applies a machine learning approach to the extracted features from a combination of static and dynamic analysis techniques in order to improve the detection performance. Shabtai et al ~\cite{Shabtai2012} presented a dynamic framework called Andromaly which applies several different machine learning algorithms, including random forest, naive Bayes, multilayer perceptron, Bayes net, logistic, and J48 to classify the Android applications. However, they assessed their performances on four self-written malware applications. MADAM ~\cite{MADAM} is also a dynamic analysis framework that uses machine learning to classify Android apps. MADAM extracted 13 features at the user and kernel level. However, their experiments were only performed on an emulator with a small dataset. Crowdroid \cite{Crowdroid} is a cloud-based machine learning framework for Android malware detection. Crowdroid features were collected based on Strace from only two self-written malware samples. Most of these previous works utilize dynamic features extracted from emulator-based analysis. By contrast, in this paper our work is based on dynamically extracted features from real device and we perform a comparative analysis between emulator and phone based machine learning approaches.

BareDroid ~\cite{Mutti2015} proposed a system designed to make bare-metal analysis of Android applications feasible. It presented analysis with malware samples from Android.HeHe ~\cite{Android.HeHe}, OBAD ~\cite{OBAD}, and Android\_Pincer.A ~\cite{AndroidPincer} families. Their work highlighted the anti-emulator capabilities of malware which can be solved by using real devices. Glassbox ~\cite{Irolla2016} also presented a dynamic analysis platform for analysing Android malware on real devices. However, unlike the work presented in this paper, these studies have not addressed machine learning based detection on real devices. Different from the previous studies, this paper presents a comparative analysis of machine learning based detection between real devices and emulators and investigates the effectiveness of run-time feature extraction in both environments.

\section{Conclusions}
In this paper we presented an investigation of machine learning based malware detection using dynamic analysis on real Android devices. We implemented a tool to automatically extract dynamic features from Android phones and through several experiments we performed a comparative analysis of emulator based vs. device based detection by means of Random Forest, Naive Bayes, Multilayer Perceptron, Simple Logistics, J48 decision tree, PART, and SVM (linear) algorithms. Our experiments showed that several features were extracted more effectively from the phone than the emulator using the same dataset. Furthermore, 23.8\% more apps were fully analyzed on the phone compared to emulator. This shows that for more efficient analysis the phone is definitely a better environment as far more apps crash when being analysed on the emulator. The results of our phone-based analysis obtained up to 0.926 F-measure and 93.1\% TPR and 92\% FPR with the Random Forest classifier and in general, phone-based results were better than emulator based results. Thus we conclude that as an incentive to reduce the impact of malware anti-emulation and environmental shortcomings of emulators which affect analysis efficiency, it is important to develop more effective machine learning device based detection solutions. Hence future work will aim to investigate more effective, larger scale device based machine learning solutions using larger sample datasets. Future work could also investigate alternative set of dynamic features to those utilized in this study.

%
\bibliographystyle{abbrv}
\bibliography{bibliography}  
%
%
\end{document}